\documentstyle[12pt,epsf]{article}

\catcode`\@=11

\begin{document}
\begin{titlepage}
\samepage{
\setcounter{page}{1}
\rightline{MADPH-98-1092}
\rightline{FERMILAB-PUB-98/364}
\rightline{hep-ph/9811350}
\vfill
\begin{center}
{\Large \bf Kaluza-Klein States  from Large Extra Dimensions\\}
\vspace{.3in}
 {\large Tao Han$^{(a)}$, 
Joseph D. Lykken$^{(b)}$ and Ren-Jie Zhang$^{(a)}$\\}
\vspace{.25in}
{\it $^{(a)}$Department of Physics, University of Wisconsin, 
     Madison, WI 53706\\[2mm]}
{\it $^{(b)}$Theory Group, Fermi National Accelerator Laboratory,
     Batavia, IL 60510\\}
\end{center}
\vfill
\vfill
\begin{abstract}
{\rm We consider the novel Kaluza-Klein (KK) scenario where
gravity propagates in the $4+n$ dimensional bulk of
spacetime, while gauge and matter fields are confined
to the $3+1$ dimensional world-volume of a brane configuration.
For simplicity we assume
compactification of the extra $n$ dimensions on a torus with a
common scale $R$,
and identify the massive KK states in the four-dimensional spacetime. 
For a given KK level
$\vec{n}$ there are one spin-2 state, $(n-1)$ spin-1 states and 
$n(n-1)/2$ spin-0 states, all mass-degenerate. 
We construct
the effective interactions between these KK states and ordinary
matter fields (fermions, gauge bosons and scalars).
We find that the spin-1 states decouple and that the spin-0 states
only couple through the dilaton mode.  
We then derive the interacting Lagrangian for the KK states and 
Standard Model fields, and present the complete Feynman rules. 
We discuss some low energy phenomenology for these new interactions
for the case when $1/R$ is small compared to the electroweak scale,
and the ultraviolet cutoff of the effective KK theory is on the
order of 1 TeV. }
\end{abstract}
\vfill}
\centerline{\it Published in Phys. Rev. D59, 105006 (1999).}
\end{titlepage}

\newcommand{\lambdab}{{\overline\lambda}}
\newcommand{\sigmab}{{\overline\sigma}}
\newcommand{\cM}{{\cal M}}


\catcode`@=11
\long\def\@caption#1[#2]#3{\par\addcontentsline{\csname
  ext@#1\endcsname}{#1}{\protect\numberline{\csname
  the#1\endcsname}{\ignorespaces #2}}\begingroup
    \small
    \@parboxrestore
    \@makecaption{\csname fnum@#1\endcsname}{\ignorespaces #3}\par
  \endgroup}
\catcode`@=12


\newcommand{ \slashchar }[1]{\setbox0=\hbox{$#1$}   
   \dimen0=\wd0                                     
   \setbox1=\hbox{/} \dimen1=\wd1                   
   \ifdim\dimen0>\dimen1                            
      \rlap{\hbox to \dimen0{\hfil/\hfil}}          
      #1                                            
   \else                                            
      \rlap{\hbox to \dimen1{\hfil$#1$\hfil}}       
      /                                             
   \fi}                                             %


\newcommand{\hmu}{{\hat\mu}}
\newcommand{\hnu}{{\hat\nu}}
\newcommand{\hrho}{{\hat\rho}}
\newcommand{\hh}{{\hat{h}}}
\newcommand{\hg}{{\hat{g}}}
\newcommand{\hk}{{\hat\kappa}}
\newcommand{\tA}{{\widetilde{A}}}
\newcommand{\tP}{{\widetilde{P}}}
\newcommand{\tF}{{\widetilde{F}}}
\newcommand{\th}{{\widetilde{h}}}
\newcommand{\tp}{{\widetilde\phi}}
\newcommand{\tchi}{{\widetilde\chi}}
\newcommand{\te}{{\widetilde\eta}}
\newcommand{\vn}{{\vec{n}}}
\newcommand{\vm}{{\vec{m}}}
\newcommand{\gsim}{\lower.7ex\hbox{$\;\stackrel{\textstyle>}{\sim}\;$}}
\newcommand{\lsim}{\lower.7ex\hbox{$\;\stackrel{\textstyle<}{\sim}\;$}}


\section{Introduction} \label{sec:intro}

Kaluza-Klein (KK) reduction \cite{rep} has always been
an important ingredient
in our attempts to relate $d=4$ physics to $d=10$ superstrings,
as well as to $d=11$ supergravity, 
which is now recognized as the
low energy effective description of $d=11$ M-theory \cite{HW}.
It has become clear, however, that a much more
general notion of Kaluza-Klein reduction is applicable in
certain regions of the moduli space of consistent
superstring/M theory vacua. This occurs when various matter and/or
gauge fields are confined to heavy solitonic membranes.
These recent developments \cite{polchinski}
in superstring theory have led to a
radical rethinking of the possibilities for new particles and
dynamics arising from extra compactified spatial dimensions
\cite{antben}-\cite{list2}.

To appreciate this radical change of view, it is useful to review
the conventional Kaluza-Klein scenario \cite{KK}. One begins with a
$d=4+n$ dimensional spacetime action, describing a coupled
gravity+gauge+matter system. Since field theories of gravity
are poorly behaved in the ultraviolet, Kaluza-Klein formulations
should be generically regarded as {\it effective} actions, with
an implicit or explicit ultraviolet cutoff $\Lambda$.
One expands this theory
around a vacuum metric which is the product of $d=4$ Minkowski space
with some $n$-dimensional compact manifold, obtained by stationarizing
this higher-dimensional effective action. For consistency, the
characteristic length scales $R_i$ of the compact manifold should be
larger than $1/\Lambda$. In the shifted vacuum all fields are
expanded in normal modes of the $n$-dimensional compact manifold;
the coefficients of this harmonic expansion are conventional
$d=4$ fields. This Kaluza-Klein reduction results in an effective $d=4$
theory of gravity+gauge+massless matter coupled to towers of massive
Kaluza-Klein states, where the massive spectrum is cutoff at
the high scale $\Lambda$.

Letting $E$ denote the energy scale of some
experiment, and assuming for simplicity that the compactification
scales $R_i\sim R$ are all roughly equal, one can distinguish three
general phenomenological regimes:

\begin{enumerate}
\item $E\ll 1/R \lsim \Lambda$. This case is relevant to
compactifications of the weakly coupled heterotic string,
with $\Lambda$ equal to the string scale, approximately
$10^{18}$ GeV. In such a case massive Kaluza-Klein modes only
impact low energy physics indirectly, through threshold effects
on couplings at the high scale.

\item $E< 1/R \ll \Lambda$. This encompasses
Kaluza-Klein scenarios where the cutoff scale $\Lambda$ is
still very high, but some dynamics fixes $1/R$
to a much lower scale, perhaps as low as a few TeV.
In this case a very large number $\sim (\Lambda R)^n$
of massive KK states are integrated out in evolving the
effective action from the high scale to the low scale.
Thus, although the couplings of individual massive KK
modes are Planck suppressed, they may contribute non-negligible
higher dimension operators
to the effective low energy theory \cite{antben}.
Furthermore, they may have strong effects on the running of the
renormalizable Standard Model (SM) couplings \cite{DDG} above the scale
$1/R$.
 
\item $1/R \ll E < \Lambda$. In this case
a large number $\sim (E\,R)^n$ of massive
KK states are kinematically accessible.
This effectively makes physics look $(4+n)$-dimensional
at the energy scale $E \gg 1/R$.
There are
severe constraints from experiment on such scenarios.
We know that
$d=4$ electrodynamics can be distinguished in collider
experiments from $d=4+n$
electrodynamics down to very short length scales.
There is also a strong bound from the non-observance
of mirror copies of the Standard Model chiral fermions.
Consider for example $d=5$ fermions, which are 
pseudo-Majorana and have four (on-shell) real degrees
of freedom. When dimensionally reduced to $d=4$, each splits into
two Weyl fermions with opposite chirality but the same gauge
group representation; therefore one expects 
mirror fermions with masses $\lsim 1/R$. 
\end{enumerate}

Recently it was observed \cite{dimopoulos}
that this last case can be phenomenologically viable if we
assume that the fields of the Standard Model are confined
to a three-dimensional membrane or intersection of membranes in
the larger dimensional space. Assuming further that the scale
of the membrane tension is on the order of the cutoff $\Lambda$
or larger, the resulting effective theory consists of 
$(3+1)$-dimensional Standard Model fields coupled to $4+n$ gravity
and, perhaps, other $(4+n)$-dimensional ``bulk'' fields.
With these assumptions the phenomenological constraints from
gravity experiments, collider physics, and astrophysics are
much weaker \cite{dimopoulos},
allowing $1/R$ scales as low as $10^{-4}$ eV ($\sim 1$ mm$^{-1}$), for cutoff
scales $\Lambda$ in the range $1-10$ TeV.

In superstring theory there are regions of moduli space
where compactification radii become large while the string
coupling, gauge couplings, and Newton's constant remain
fixed \cite{witten,lykken}.
The scale of these large extra dimensions is related to the
string scale $M_S$:
\begin{equation}
{1\over G_N} \sim M_S^{n+2}\,R^n\ ,
\label{mas}
\end{equation}
where $G_N$ is the Newton constant. 
Roughly speaking, $M_S$ plays the role of the ultraviolet
cutoff $\Lambda$. This reproduces the relationship of scales
assumed in the scenario just described.

It has also been shown in superstring theory that it is
possible to obtain $d=4$ N=1 supersymmetric chiral gauge
theories confined to the world-volumes of stable configurations
of intersecting D-branes \cite{chiral}. The regions of string moduli
space where such configurations have a perturbative description
is not necessarily incompatible with the region where large
extra dimensions may occur.
Thus within our current knowledge (or ignorance) of superstrings 
it is not implausible to imagine that the Standard Model is
confined to a brane configuration \cite{past,sundrum},
while large compactified
dimensions are probed only by gravity and other bulk
fields \cite{dimopoulos,tye}.

In this paper we will consider the simplest case where
gravity is the only $d=4+n$ bulk field. The couplings
of gravity to $d=4$ gauge and matter fields are completely
fixed by general coordinate invariance in the $d=4+n$
spacetime and the $d=4$ world-volume. This allows us to
deduce the complete Feynman rules for the couplings of
Standard Model particles to the massive KK states.
The low energy phenomenology is then calculable modulo
the details of how to treat the cutoff $\Lambda$, which
truncates the KK mode sums.

In the following, we will use hatted letters to denote 
the $(4+n)$-dimensional quantities,
{\it e.g.}, $\hg_{\hmu\hnu}$ denotes the metric tensor in $d=4+n$.
Un-hatted Greek letters ($\mu, \nu, \cdots$),  
Roman letters from the beginning ($a, b,\cdots$)
and in the middle ($i, j, \cdots$) of the alphabet will be used to
label four-dimensional Einstein, Lorentz and
(the compactified) $n$-dimensional indices 
respectively. Repeated indices are summed.  
Our convention for the signature is $(+,-,-,\cdots)$.

The rest of the paper is organized as follows: In Section 2, 
we compactify $d=4+n$ gravity on an $n$-dimensional torus $T^n$
and perform a mode expansion. A torus compactification is perhaps
not realistic, since the bulk fields which we are ignoring are
potential sources of $n$-dimensional curvature, as are the branes
themselves.
However the torus has the great advantage of conceptual and
calculational simplicity.  
We find that the massive KK modes 
have a simple physical interpretation. For each KK level, 
there are one massive spin-2, $(n-1)$ massive
spin-1 and $n(n-1)/2$ massive spin-0 particles. We
find the general
form for the interactions between matter
(scalars, gauge bosons and fermions) and the massive
KK states.
In Section 3, we examine a few physical processes involving
the KK states. We calculate their decay widths to the
light SM particles; this could have important cosmological consequences.
We then construct effective four-fermion and ${\bar f}fVV$
interactions; this provides a useful formalism for studying some
high energy processes. We next study the process 
$e^+e^-\rightarrow\gamma +KK$, where $KK$ are spin-0 and 2 massive
KK states.  In the final example,
we calculate the one-loop corrections to the scalar boson
masses due to virtual KK states; we find that the corrections are
proportional to the scalar mass, instead of the ultraviolet cutoff
$M_S$. Section 4 is reserved for the
discussions and conclusions. 
We list some useful formulae in two appendices.
In Appendix A, we present the propagators and polarizations for the 
physical KK states, and show
the complete leading-order (${\cal O}(\kappa)$)
vertex Feynman rules. In Appendix B, we discuss 
the summation over KK states which appears in many physical processes.

\section{General Formalism}

\subsection{Decomposition of the Massive KK States}

The starting point for our analysis is the linearized gravity Lagrangian, 
{\it i.e.}, the Fierz-Pauli Lagrangian \cite{veltman}: 
\begin{equation}
{1\over\hk^2} \sqrt{|{\hat g}|} {\hat R}\ =\ 
{1\over4}\biggl(
\partial^\hmu\hh^{\hnu\hrho} \partial_\hmu\hh_{\hnu\hrho} 
-\partial^\hmu\hh \partial_\hmu\hh
-2\hh^{\hmu} \hh_{\hmu}
+2\hh^{\hmu} \partial_\hmu\hh\Biggr) + {\cal O} (\hk)\ ,
\label{FP}
\end{equation}
where $\hh\equiv\hh^\hmu_{~\hmu}, \hh_\hnu\equiv\partial^\hmu\hh_{\hmu\hnu}$ 
and we have used $\hat{g}_{\hmu\hnu}=\eta_{\hmu\hnu}+\hk \hh_{\hmu\hnu}$,
$\hat\kappa^2=16\pi G_N^{(4+n)}$, with $G_N^{(4+n)}$ the Newton constant
in $d=4+n$.
This Lagrangian is invariant under the general coordinate
transformation
\begin{equation}
\delta\hh_{\hmu\hnu}\ =\ \partial_\hmu\zeta_\hnu + \partial_\hnu\zeta_\hmu\ 
.
\label{trans2}
\end{equation}

After imposing the de Donder gauge condition\footnote{Here we choose the 
gauge condition for the sake of clarity; the definitions of physical
fields in Eq.~(\ref{redef}) do not depend on the gauge choice.}
\begin{equation}
\partial^\hmu (\hh_{\hmu\hnu}-{1\over2}\eta_{\hmu\hnu}\hh) =0 \ ,
\label{gauge}
\end{equation}
the equation of motion is the d'Alembert equation
\begin{equation}
\Box_{(4+n)}\ (\hh_{\hmu\hnu}-{1\over2}\eta_{\hmu\hnu}\hh)\ =\ 0\ .
\label{hareom}
\end{equation}
The gauge condition, along with the tracelessness condition
$\hh^\hmu_{~\hmu} = 0$, and the residual
general coordinate transformation Eq.~(\ref{trans2}), with the gauge
parameter satisfying $\Box_{(4+n)} \zeta_\hmu=0$,
fixes all but the $(2+n)(3+n)/2-1$ physical degrees of freedom for
a massless graviton in $4+n$ dimensions.

Now we proceed to perform the KK reduction of the Fierz-Pauli
Lagrangian to $d=4$. We shall assume 
\begin{equation}
\hh_{\hmu\hnu}\ =\ 
V_n^{-1/2}\left(\begin{array}{cc}
h_{\mu\nu}+\eta_{\mu\nu}\phi & A_{\mu i}\\
A_{\nu j}   &  2 \phi_{ij}
\end{array}\right)\ ,
\label{gmunu}
\end{equation}
where $V_n$ is the volume of the $d=n$ compactified space,
$\phi\equiv\phi_{ii}$, $\mu, \nu=0,1,2,3$ and $i, j=5,6,\cdots ,4+n$,
and the $\eta_{\mu\nu}\phi$ term in the $(11)$-entry is a Weyl rescaling. 
These fields are compactified on an $n$-dimensional torus $T^n$ and
have the following mode expansions:
\begin{eqnarray}
&& h_{\mu\nu}(x,y)\ =\ \sum_{\vec n} h_{\mu\nu}^{\vec{n}}(x)\ 
\exp\left(i {2\pi \vec{n}\cdot\vec{y}\over R}\right)\ ,\\
&& A_{\mu i}(x,y)\ =\ \sum_{\vec n} A_{\mu i}^{\vec{n}}(x)\ 
\exp\left(i {2\pi \vec{n}\cdot\vec{y}\over R}\right)\ ,\\ 
&& \phi_{ij}(x,y)\ =\ \sum_{\vec n} \phi_{ij}^{\vec{n}}(x)\ 
\exp\left(i {2\pi \vec{n}\cdot\vec{y}\over R}\right)\ , \qquad
{\vec n} = \{n_1,n_2,\cdots,n_n\}\ ,\label{mode}
\end{eqnarray}
where the modes of $\vec{n}\neq 0$ are 
the KK states, and all the compactification radii are assumed 
to be the same as $R/2\pi$. The generalization 
to an asymmetric torus with different radii is straightforward.
From the transformation properties under the general
coordinate transformation $\zeta_{\hmu}=\{\zeta_\mu, \zeta_i\}$,
it should be clear that the zero modes, $\vn=\vec{0}$,  correspond
to the massless graviton, U(1) gauge bosons and scalars in $d=4$.

The KK modes satisfy the following equation of motions, 
from Eq.~(\ref{hareom}),
\begin{eqnarray}
&&
(\Box + m^2_{\vec{n}})\ (h^{\vec{n}}_{\mu\nu}-{1\over2}\eta_{\mu\nu}h^\vn)
\ =\ 0\ ,~~~~
(\Box + m^2_{\vec{n}})\ A^{\vec{n}}_{\mu i}\ =\ 0\ ,\nonumber\\
&&(\Box + m^2_{\vec{n}})\ \phi_{ij}^{\vec{n}}\ =\ 0\ ,
\qquad {\rm where\ }
m^2_{\vec{n}}\ =\ {4\pi^2 {\vec{n}}^2\over R^2}\ ,
\label{eom}
\end{eqnarray}
and $\Box$ is the four-dimensional d'Alembert operator.
The gauge condition in Eq.~(\ref{gauge}) reduces to the following two
equations
\begin{eqnarray}
&&\partial^\mu h^{\vec{n}}_{\mu\nu}-{1\over2}\partial_\nu h^{\vec{n}}
+ i {2\pi n_i\over R} A^{\vec{n}}_{\nu i}\ =\ 0\ ,\label{2a}\\
&&\partial^\mu A_{\mu i}^{\vec{n}} + i  {4\pi n_j\over R}\phi^{\vec{n}}_{ij}
+ i {\pi n_i\over R} h^{\vec{n}} + i {2\pi n_i\over R}\phi^{\vec{n}}\ =\ 0\ .
\label{2b}
\end{eqnarray}

From Eq.~(\ref{2b}), it follows
\begin{eqnarray}
&&\phi^{\vec{n}} + {2n_i n_j\over{\vec{n}}^2}\phi_{ij}^{\vec{n}}
+ {1\over2}h^{\vec{n}} - i {n_i R\over 2\pi {\vec{n}}^2} \partial^\mu 
A_{\mu i}^{\vec{n}}\ =\ 0\ ,\label{3a}\\
&& P^\vn_{ik} (\partial^\mu A_{\mu i}^{\vec{n}} + i {4\pi n_j\over R}
\phi_{ij}^{\vec{n}})\ =\ 0\ ,\label{3b}
\end{eqnarray}
where we have defined projectors
\begin{equation}
P^\vn_{ij}\ =\ \delta_{ij} - {n_i n_j\over{\vec{n}}^2}\ ,\qquad
\widetilde{P}^\vn_{ij}\ =\ {n_i n_j\over{\vec{n}}^2}\ ,
\label{proj}
\end{equation}
they satisfy
\begin{eqnarray}
&&P_{ij}^\vn P_{jk}^\vn\ =\ P_{ik}^\vn\ ,~~
\tP_{ij}^\vn \tP_{jk}^\vn\ =\ \tP_{ik}^\vn\ ,~~
P_{ij}^\vn\tP_{jk}^\vn\ =\ 0\ ,
~~P^\vn_{ij}+\tP^\vn_{ij}\ =\ \delta_{ij}\ ,\nonumber\\
&&P_{ii}^\vn\ =\ n-1\ , ~~\tP_{ii}^\vn\ =\ 1\ ,~~
P_{ij}^\vn\ n_i\ =\ 0\ ,~~\tP_{ij}^\vn\ n_i\ =\ n_j\ .
\end{eqnarray}

We then redefine the fields 
\begin{eqnarray}
\th_{\mu\nu}^{\vec{n}}&=& 
h_{\mu\nu}^{\vec{n}} - i {n_i R\over 2\pi{\vec{n}}^2}
(\partial_\mu A_{\nu i}^{\vec{n}}+\partial_\nu A_{\mu i}^{\vec{n}})
-(P^\vn_{ij}+3 {\widetilde{P}}^\vn_{ij})
\biggl({2\over3}{\partial_{\mu}\partial_\nu\over m_\vn^2}
-{1\over3}\eta_{\mu\nu}\biggr)\phi_{ij}^{\vec{n}}\ ,\nonumber\\
\tA_{\mu i}^{\vec{n}} &=& P^\vn_{ij} (A_{\mu j}^{\vec{n}} 
- i {n_k R\over\pi{\vec{n}}^2}\partial_\mu\phi^{\vec{n}}_{jk})\ ,
\qquad~~ \tp_{ij}^{\vec{n}}\ =\ \sqrt{2}
(P^\vn_{ik} P^\vn_{jl} + a P_{ij}^\vn P_{kl}^\vn) \phi_{kl}^{\vec{n}}\ ,
\label{redef}
\end{eqnarray}
where $a$ is the solution of the equation $3(n-1)a^2+6a=1$.
This form of $\tp_{ij}^\vn$ is chosen to make its kinetic term canonical,
as will be seen in Eq.~(\ref{KKlag}).
It is obvious that tilded fields satisfy the same
equations of motion as untilded fields. Furthermore from Eqs. (\ref{2a}),
(\ref{3a}), (\ref{3b}) and (\ref{redef}), we have
\begin{eqnarray}
&&\partial^\mu \th^{\vec{n}}_{\mu\nu}
\ =\ 0\ ,\qquad \th^{\vec{n}}\ =\ 0\ ,\label{constraint1}\\
&&\partial^\mu \tA_{\mu i}^{\vec{n}}\ =\ 0\ ,
\qquad n_i \tA_{\mu i}^{\vec{n}}\ =\ 0\ ,
\qquad n_i \tp_{ij}^{\vec{n}}\ =\ 0\ .
\label{constraint2}
\end{eqnarray}
This verifies that $\th_{\mu\nu}^{\vec{n}}$ are massive spin-2 particles,
$\tA_{\mu i}^{\vec{n}}$ are $(n-1)$ massive spin-1 particles, and
$\tp_{ij}^{\vec{n}}$ are $n(n-1)/2$ massive spin-0 particles,
all with the same mass $m_\vn$.

This redefinition of fields is associated with spontaneous symmetry
breaking. 
It was shown for $n=1$ there is 
an infinite-dimensional symmetry (the loop algebra on $S^1$) 
at the Lagrangian level \cite{duff},
but it is broken by the vacuum configuration, 
$\hg_{\hmu\hnu}=\eta_{\hmu\hnu}$.
Similar to the Higgs mechanism, the massless spin-2 fields 
$h_{\mu\nu}^{\vec{n}}$ absorb the spin-1 and spin-0 fields 
at the same KK level $\vec{n}$ and become massive.
It is remarkable that this mechanism is geometrical in nature and does 
not need any scalar Higgs field. We here explicitly find the composition
for massive spin-2, 1 and 0 fields for $n\geq 2$.

One can further show that $\th_{\mu\nu}^{\vec{n}}$,
$\tA_{\mu i}^{\vec{n}}$ and $\tp_{ij}^{\vec{n}}$ are invariant under
the general coordinate transformation, which has the following
linearized form
\begin{eqnarray}
&&\delta h_{\mu\nu}^{\vec{n}}\ =\ 
\partial_\mu \zeta_\nu^{\vec{n}}+\partial_\nu \zeta_\mu^{\vec{n}}
+ i\eta_{\mu\nu}{2\pi n_i\over R}\zeta_i^{\vec{n}}\ ,\\
&&\delta A_{\mu i}^{\vec{n}}\ =\ 
-i{2\pi n_i\over R} \zeta_\mu^{\vec{n}}+\partial_\mu \zeta_i^{\vec{n}}\ ,\\
&&\delta \phi_{ij}^{\vec{n}}\ =\ 
-i{\pi n_i\over R}\zeta_j^{\vec{n}}-i{\pi n_j\over R}\zeta_i^{\vec{n}}\ ,
\end{eqnarray}
where we have assumed the transformation parameters 
$\zeta_\mu^\vn, \zeta^\vn_i$ to have the same mode expansion as 
in Eq.~(\ref{mode}). 

We should note that 
the field redefinition in Eq.~(\ref{redef}) does not depend on 
the particular gauge choice. To see this, 
we rewrite the Lagrangian in Eq.~(\ref{FP})
without imposing the de Donder gauge.
For the zero modes, it simply follows from Eq.~(\ref{gmunu}),
\begin{eqnarray}
{\cal L}^{\vec{0}}&=& 
{1\over4}\biggl(
\partial^\mu h^{\nu\rho} \partial_\mu h_{\nu\rho} 
-\partial^\mu h \partial_\mu h
-2 h^{\mu} h_{\mu} + 2 h^{\mu} \partial_\mu h\Biggr)\nonumber\\ 
&&-\sum_{i=1}^n {1\over4} F_i^{\mu\nu} F_{\mu\nu i}
+{1\over2}\partial^\mu \phi\partial_\mu \phi
+\sum_{(ij)=1}^{n(n+1)/2}\partial^\mu\phi_{ij}\partial_\mu\phi_{ij}\ ,
\end{eqnarray}
where $F_{\mu\nu i} =\partial_\mu A_{\nu i}-\partial_\nu A_{\mu i}$,
and $(ij)$ is viewed as one index with symmetrization of $i$ and $j$.
We see it indeed describes massless graviton, vectors and scalars.

The Lagrangian for the massive KK modes can be rewritten in terms 
of the tilded fields according to Eq.~(\ref{redef}).
After a tedious calculation, we find
\begin{eqnarray}
{\cal L}^{\vec{n}}+{\cal L}^{\vec{-n}}&=& 
{1\over2}\biggl(
\partial^\mu\th^{\nu\rho,\vn} \partial_\mu\th_{\nu\rho}^{-\vn} 
-\partial^\mu\th^\vn \partial_\mu\th^{-\vn}
-2\th^{\mu,\vn}\th_{\mu}^{-\vn} 
+ \th^{\mu,\vn}\partial_\mu\th^{-\vn}
+ \th^{\mu,-\vn}\partial_\mu\th^{\vn}\nonumber\\
&&\qquad -m^2_\vn\th^{\mu\nu,\vn}\th_{\mu\nu}^{-\vn}
+m^2_\vn\th^{\vn}\th^{-\vn}\Biggr)
+\sum_{i=1}^n
(-{1\over2}\tF_i^{\mu\nu,\vn}\tF_{\mu\nu i}^{-\vn}
+ m^2_\vn\tA_i^{\mu,\vn}\tA_{\mu i}^{-\vn})\nonumber\\
&&
+\sum_{(ij)=1}^{n(n+1)/2}
(\partial^\mu\tp^\vn_{ij}\partial_\mu\tp^{-\vn}_{ij}
-m^2_\vn\tp^\vn_{ij}\tp^{-\vn}_{ij})\ ,
\label{KKlag}
\end{eqnarray}
the fields $\tA_{\mu i}^\vn$ and $\tp_{ij}^\vn$ are subject
to the constraints in Eq.~(\ref{constraint2}). 

The equation
of motion of $\th_{\mu\nu}^\vn$
from Eq.~(\ref{KKlag}) is the Fierz-Pauli equation
for massive spin-2 particles 
\begin{equation}
\partial^\mu\partial_\mu\tchi^\vn_{\nu\rho}
-\partial_\nu\tchi^\vn_{\rho}-\partial_\rho\tchi^\vn_{\nu}
+\partial^\mu\tchi^\vn_{\mu}\eta_{\nu\rho}+m_\vn^2(\th_{\nu\rho}^\vn
-\eta_{\nu\rho}\th^\vn)
\ =\ 0\ ,
\end{equation}
where 
\begin{equation}
\tchi_{\mu\nu}^\vn\ =\ \th^\vn_{\mu\nu}-{1\over2}\th^\vn\eta_{\mu\nu}\ ,
~~\tchi_\mu^\vn\ =\ \partial^\nu\tchi_{\mu\nu}^\vn\ ,
\end{equation}
and $\tA^\vn_{\mu i}$ and $\tp^\vn_{ij}$ satisfy
\begin{equation}
\partial^\mu\tF_{\mu\nu i}^\vn + m_\vn^2 \tA_{\nu i}^\vn\ =\ 0\ ,
\qquad
(\Box+m^2_\vn) \tp^\vn_{ij}\ =\ 0\ ,
\end{equation}
These equations can be recast into the form in 
Eqs. (\ref{eom}), (\ref{constraint1}) and (\ref{constraint2}).

The propagators and polarizations of the physical (tilded) 
fields will be given in Appendix A.1.

\subsection{Coupling of the KK States to Matter}

The basic picture for our physical world, as considered in 
this paper, is that all Standard Model
fields are confined to a four-dimensional brane world-volume.
As we showed in the previous section, from the four-dimensional 
perspective, the zero modes of the $(4+n)$-dimensional graviton
become the graviton, $n$ massless U(1) gauge bosons and
$n(n+1)/2$ massless scalar bosons, while the KK modes in each level
reorganize themselves to 
a massive spin-2 particle, $(n-1)$ massive vector bosons and
$n(n-1)/2$ massive scalar bosons.    
In the following, we will formulate the coupling of these physical
KK modes to the matter. Although these interactions only
have gravitational strength, they can be enhanced 
in the case of large size extra dimensions, 
due to the many available KK states. 

We begin with the minimal gravitational coupling of the 
general scalar $S$, vector $V$, and fermion $F$\footnote{For the fermion,
one should use the vierbein formalism, but our result in Eq.~(\ref{gen})
is still true.}, 
\begin{equation}
\int d^4x \sqrt{-\hat{g}}\ {\cal L}(\hg,S,V,F)\ ,
\label{coup}
\end{equation}
where $\hat{g}$ is the induced metric in $d=4$, $\hg_{\mu\nu}
=\eta_{\mu\nu} + \kappa (h_{\mu\nu} + \eta_{\mu\nu} \phi)$, 
$\phi\equiv\phi_{ii}$. The $d=4$
Newton constant $\kappa=\sqrt{16\pi G_N}$ 
is related to $\hat{\kappa}$ by $\kappa=V_n^{-1/2}\hat\kappa$,
where $V_n=R^n$ for the torus $T^n$.

The ${\cal O}(\kappa)$ term of Eq.~(\ref{coup}) can be easily shown to be
\begin{equation}
-{\kappa\over 2} \int d^4 x (h^{\mu\nu} T_{\mu\nu}
+ \phi T^\mu_{~\mu})\ ,
\end{equation}
where 
\begin{equation}
T_{\mu\nu}(S,V,F)\ =\ \biggl(-{\eta_{\mu\nu}}{\cal L}
+2 {\delta{\cal L}\over\delta\hg^{\mu\nu}} 
\biggr)|_{\hg=\eta}\ , 
\end{equation}
and we have used 
\begin{equation}
\sqrt{-\hg}\ =\ 1+{\kappa\over2}h + 2\kappa\phi\ , 
\ \ 
\hg^{\mu\nu}\ =\ \eta^{\mu\nu}-\kappa h^{\mu\nu}-\kappa\eta^{\mu\nu}\phi\ .
\end{equation}
For the KK modes, we should replace 
$h^{\vec{n}}_{\mu\nu}$ and $\phi^\vn$ 
by the physical fields $\th^{\vec{n}}_{\mu\nu}$ and $\tp^\vn$
according to Eq.~(\ref{redef}). Using  
\begin{equation}
P^\vn_{ij}\phi^{\vec{n}}_{ij}\ =\ 
{3\omega\over2}\tp^{\vec{n}}\ ,
\end{equation}
where $\tp^{\vec{n}}\ \equiv\ \tp^{\vec{n}}_{ii}$, 
$\omega=\sqrt{2\over3(n+2)}$, 
and the conservation of the energy-momentum tensor, we obtain
\begin{equation}
-{\kappa\over 2}
\sum_\vn \int d^4 x (\th^{\mu\nu,\vec{n}} T_{\mu\nu} 
+ \omega\tp^{\vec{n}} T^\mu_{~\mu})\ .
\label{gen}
\end{equation}
It is remarkable that the vector KK modes $\tA_{\mu i}^{\vec{n}}$
decouple and the scalar KK modes $\tp_{ij}^{\vec{n}}$ only couple through 
their trace $\tp^{\vec{n}}$, the dilaton mode.

We now present the Lagrangian to the order of ${\cal O}(\kappa)$; 
a complete list of vertex functions will be given in Appendix A.2.

\subsubsection{Coupling to Scalar Bosons}

For a general complex scalar field $\Phi$, we have
the conserved energy-momentum tensor
\begin{equation}
T^{\rm S}_{\mu\nu}\ =\ -\eta_{\mu\nu}D^\rho\Phi^\dagger D_\rho\Phi
        +\eta_{\mu\nu}m^2_\Phi\Phi^\dagger\Phi
	+D_\mu\Phi^\dagger D_\nu\Phi
	+D_\nu\Phi^\dagger D_\mu\Phi\ ,
\end{equation}
where the gauge covariant derivative is defined as
\begin{equation}
D_\mu = \partial_\mu + i g A^a_\mu T^a,
\end{equation}
with $g$ the gauge coupling,
$A^a_\mu$ the gauge fields and $T^a$ the Lie algebra
generators. The gauge-invariant Lagrangian for a level-$\vn$ KK 
state coupled to the scalar bosons is
\begin{eqnarray}
\kappa^{-1}{\cal L}^{\vec{n}}_{\rm S}(\kappa) &=& 
  -(\th^{\mu\nu,\vec{n}}
  -{1\over2}\eta^{\mu\nu}\th^\vn) D_\mu\Phi^\dagger D_\nu\Phi 
  -{1\over2}\th^\vn m_\Phi^2 \Phi^\dagger\Phi\nonumber\\
&&+\omega\tp^\vn (D^\mu\Phi^\dagger D_\mu\Phi 
  -2m_\Phi^2\Phi^\dagger\Phi)\ .
\end{eqnarray}
From this, one finds the Feynman rules 
for KK-$\Phi\Phi$ vertices as well as the contact interactions
of KK-$\Phi\Phi$ with additional 
gauge bosons. They are listed in Appendix A.2.

\subsubsection{Coupling to Gauge Bosons}

The conserved energy-momentum tensor for a gauge vector boson is
\begin{eqnarray}
T^{\rm V}_{\mu\nu} &=& \eta_{\mu\nu}
\biggl({1\over4} F^{\rho\sigma}
F_{\rho\sigma}-{m^2_A\over2} A^\rho A_\rho\biggr)
- \biggl(F_\mu^{~\rho} F_{\nu\rho} 
- m_A^2 A_\mu A_\nu\biggr)\nonumber\\
&&
-{1\over\xi}\eta_{\mu\nu}\biggl(\partial^\rho\partial^\sigma A_\sigma A_\rho
+{1\over2}(\partial^\rho A_\rho)^2\biggr)
+{1\over\xi}(\partial_\mu\partial^\rho A_\rho A_\nu+
\partial_\nu\partial^\rho A_\rho A_\mu)\ ,
\end{eqnarray}
where the $\xi$-dependent terms correspond to adding a gauge-fixing
term $-(\partial^\mu A_\mu -\Gamma^{\mu\nu}_{~~\nu} A_\mu)^2/2\xi$, 
with $\Gamma^{\mu\nu}_{~~\nu}=\eta^{\nu\rho}\Gamma^\mu_{~\nu\rho}$ 
the Christoffel symbol (affine connection).
The Lagrangian for a level-$\vn$ KK state coupled to the
gauge bosons is
\begin{eqnarray}
\kappa^{-1}{\cal L}^{\vec{n}}_{\rm V} (\kappa) 
&=& -{1\over8}(\th^\vn\eta^{\mu\nu}-4\th^{\mu\nu,\vn})
F_\mu^{~\rho}F_{\nu\rho}+{1\over4}(\th^\vn\eta^{\mu\nu}-2\th^{\mu\nu,\vn})
m^2_A A_\mu A_\nu\nonumber\\
&&
+{\th^\vn\over2\xi}\biggl(\partial^\rho\partial^\sigma A_\sigma A_\rho
+{1\over2}(\partial^\rho A_\rho)^2\biggr)
-{\th^{\mu\nu,\vn}\over\xi}\partial_\mu\partial^\rho A_\rho A_\nu\nonumber\\
&&+{\omega\over2} m_A^2 \tp^\vn A^\mu A_\mu
-{\omega\over\xi}\partial^\mu\tp^\vn\partial^\nu A_\nu A_\mu\ .
\end{eqnarray}
The corresponding Feynman rules for three-point
KK-$AA$ vertices as well as the contact
interactions of KK-$AAA$ and KK-$AAAA$ are given in Appendix A.2.

\subsubsection{Coupling to Fermions}

To describe a fermion in the gravitation theory, one 
needs to use the vierbein formalism. The fermion Lagrangian is
\begin{equation}
{\cal L}_{\rm F}\ =\  e {\overline\psi}
(i\gamma^\mu {\cal D}_\mu - m_\psi)\psi\ ,
\end{equation} 
where $e={\rm det}(e_\mu^{~a})$, $e_\mu^{~a}e_\nu^{~b}\eta_{ab}
=g_{\mu\nu}$, $\gamma^\mu=e^\mu_{~a}\gamma^a$,
and $a, b$ are Lorentz indices. The covariant
derivative on the fermion field is defined by
\begin{equation}
{\cal D}_\mu \psi\ =\ (D_\mu + {1\over2}\omega_{\mu}^{ab}\sigma_{ab})\psi\ ,
\end{equation}
where $\sigma_{ab} = {1\over4}[\gamma_a,\gamma_b]$. In the 
absence of a spin-3/2 field, the spin connection $\omega_{\mu}^{ab}$
can be solved in terms
of the vierbein, 
\begin{equation}
\omega_{\mu ab}\ =\ 
{1\over2}(\partial_\mu e_{b\nu}-\partial_\nu e_{b\mu})e_a^{~\nu} 
-{1\over2}(\partial_\mu e_{a\nu}-\partial_\nu e_{a\mu})e_b^{~\nu} 
-{1\over2}e_a^{~\rho}e_b^{~\sigma}
(\partial_\rho e_{c\sigma}-\partial_\sigma e_{c\rho})e^c_{~\mu}\ .
\end{equation}

We find the conserved energy-momentum tensor
\begin{eqnarray}
T_{\mu\nu}^{\rm F} &=& -\eta_{\mu\nu}
(\overline{\psi}i\gamma^\rho D_\rho\psi-m_\psi\overline{\psi}\psi)
+{1\over2}\overline{\psi}i\gamma_\mu D_\nu\psi
+{1\over2}\overline{\psi}i\gamma_\nu D_\mu\psi\nonumber\\
&&+{\eta_{\mu\nu}\over2}\partial^\rho(\overline{\psi}i\gamma_\rho\psi)
-{1\over4}\partial_\mu(\overline{\psi}i\gamma_\nu\psi)
-{1\over4}\partial_\nu(\overline{\psi}i\gamma_\mu\psi)\ ,
\end{eqnarray}
where we have used the linearized vierbein 
\begin{equation}
e_\mu^{~a}\ =\ \delta_\mu^{~a}
+{\kappa\over2}(h_\mu^{~a}+\delta_\mu^{~a}\phi)\ .
\end{equation}
The Lagrangian for a level-$\vn$ KK state coupled to fermions is
\begin{eqnarray}
\kappa^{-1} {\cal L}_{\rm F}^\vn (\kappa) &=& 
{1\over2}\biggl[(\th^\vn \eta^{\mu\nu}- \th^{\mu\nu,\vn})
{\overline\psi} i\gamma_\mu D_\nu \psi 
- m_\psi \th^\vn {\overline\psi}\psi
+ {1\over2}{\overline\psi}i\gamma^\mu 
(\partial_\mu \th^\vn -\partial^\nu \th^\vn_{\mu\nu})\psi \Biggr]\nonumber\\
&&
+{3\omega\over2}\tp^\vn\overline{\psi}i\gamma^\mu D_\mu\psi 
- {2\omega} m_\psi\tp^\vn{\overline\psi}\psi
+{3\omega\over4}\partial_\mu\tp^\vn{\overline\psi}i\gamma^\mu\psi\ .
\end{eqnarray}
The Feynman rules for KK-$\psi\psi$ vertices as well as contact interactions
of KK-$\psi\psi$ with additional gauge bosons are listed in Appendix A.2.

\section{Application to Physical Processes}

We are interested in a scenario in which the experimentally
accessible energy is larger than the compactification scale
$1/R$ (from $\sim 10^{-4}$ eV to $100$ MeV for $n=2$ to 7)
but lower than the ultraviolet cutoff $\Lambda$. 
We first consider how the KK states decay to the SM particles.
We then outline some low energy phenomenology and formulate
effective amplitudes relevant to further studies at colliders.
Finally, we evaluate typical one-loop corrections from virtual
KK states to a scalar propagator. 
For simplicity, we will take the 
ultraviolet cutoff $\Lambda$ to be the string scale $M_S$.
More general choice of $\Lambda$ can be obtained by simple
scaling. 

\subsection{Decay of the Massive KK States}

A massive KK state may decay to a pair of SM particles,
beside its normal decay modes to massless gravitons and the lighter
KK states. Depending on its mass, it can go to 
$\gamma\gamma, f\bar f, WW,ZZ$ and $hh$. 
While the decay of an individual massive KK 
state may not be much of interest for the current high 
energy experiments since it must be gravitationally suppressed,
cosmological considerations of their lifetimes
may have significant implications for their masses and interactions.
Without speculating on the production and 
freeze-out of the KK modes at the early 
Universe with extra dimensions, we simply evaluate their decay widths
and lifetimes to SM particles.

\subsubsection{Spin-2 KK States}

We first consider a massive spin-2 KK state ($\tilde h$) 
decay to gauge bosons
\begin{equation}
\tilde h \to V V\ .
\end{equation}
It is straightforward to work out the partial decay width to
massless gauge bosons, 
\begin{equation}
\Gamma(\th\to V V)=N {\kappa^2 m_\th^3\over 160\pi},
\label{htogg}
\end{equation}
where $N=1\ (8)$ for photons (gluons). 

Due to the universal $\tilde h$ coupling to all gauge bosons, 
the two-photon mode $\tilde h \to \gamma \gamma$ is kinematically
most favored for the lower-lying KK states. 
The lifetime is estimated to be
\begin{equation}
\tau_{\gamma\gamma}\approx {5\times 10^2 \over \kappa^2 m_\th^3} 
\approx 3\times 10^{9}\ {\rm yr}\ 
\left( {100\ {\rm MeV}\over m_\th}\right)^3\ ,
\label{lifetime}
\end{equation}
where we have taken the reduced Planck mass
$M^*_{\rm pl}=\sqrt{2}\kappa^{-1}=2.4\times 10^{18}$ GeV.
It is very long-lived via this decay mode. For a KK state heavier
than the lower-lying hadrons, its lifetime via $\tilde h \to gg$ 
would be shorter
\begin{equation}
\tau_{gg} \approx 4\times 10^5\ {\rm yr}\ 
\left( {1\ {\rm GeV}\over m_\th}\right)^3\ .
\label{lifetimeg}
\end{equation}
If kinematically allowed, the KK mode can decay to massive
gauge bosons and the decay width is
\begin{equation}
\Gamma(\th\to V V)=\delta {\kappa^2 m_\th^3\over 80\pi}
(1-4r_V)^{1/2}({13\over 12} +{14\over 39} r_V + {4\over 13}r^2_V),
\label{htovv}
\end{equation}
where $\delta=1/2$ for identical particles.
Here and henceforth, we will use a notation for 
the mass ratio $r_i=m_i^2/m^2_\th$ or $m_i^2/m^2_\tp$.
The lifetime through this decay channel is 
\begin{equation}
\tau_{VV}^{}\approx {3\times 10^2 \over \kappa^2 m_\th^3} 
\approx  15\ {\rm yr}\ 
\left({100\ {\rm GeV}\over m_\th}\right)^3\ .
\label{lifetimev}
\end{equation}

The other decay channel goes through fermions,
\begin{equation}
\tilde h \to f \bar f.
\end{equation}
The decay width is
\begin{equation}
\Gamma(\th\to f \bar f)= N_c\ {\kappa^2 m_\th^3\over 320\pi}
(1-4r_f)^{3/2}(1 +{8\over 3} r_f),
\end{equation}
where the color factor $N_c$ is three for the quark pair mode.
The lifetime for this channel is
of the same order of magnitude as that of Eq.~(\ref{lifetime}).

Finally, the decay width to a pair of Higgs bosons is
\begin{equation}
\Gamma(\th\to H \overline H)= {\kappa^2 m_\th^3\over 960\pi}
(1-4r_H^{})^{5/2}.
\end{equation}
We notice the threshold effects for the above three modes
as $S$, $P$ and $D$ waves.

\subsubsection{Spin-0 KK States}

The spin-0 KK state ($\tp$) couplings to massless gauge bosons
vanish at tree level, so that a $\tp$ does not decay to photons
nor to gluons at the leading order. 
If kinematically allowed, a massive 
$\tp$ can decay to massive gauge bosons
\begin{equation}
\tp \to V V\ .
\end{equation}
The partial decay width is calculated to be
\begin{equation}
\Gamma(\tp \to V V)={\delta\over n+2}\ {\kappa^2 m_\tp^3\over 48\pi}
(1-4r_V)^{1/2}(1 -4 r_V + 12 r^2_V),
\end{equation}
where, again, $\delta=1/2$ for identical particles.
The lifetime based on this decay channel is about the same
order of magnitude as that of Eq.~(\ref{lifetimev}).

On the other hand, a light $\tp$ can still decay to a 
pair of light fermions
\begin{equation}
\tp \to f \bar f.
\end{equation}
The decay width is given by
\begin{equation}
\Gamma(\tp \to f \bar f)= {N_c\over n+2}\ {\kappa^2 m_f^2 m_\tp\over 24\pi}
(1-4r_f)^{1/2}(1 -2 r_f).
\end{equation}
The width for this channel is rather different from $\th$ decay,
being proportional linearly to $m_\tp$ and quadratically to $m_f$.
This is because of the fermion spin-flip interactions by a scalar.
The lifetime of $\tp$ for this channel is estimated to be
\begin{equation}
\tau\approx {3\times 10^2 \over \kappa^2 m_f^2 m_\tp} 
\approx  2\times 10^{10}\ {\rm yr}\ 
{(100\ {\rm MeV})^3\over m_f^2 m_\tp}\ .
\label{lifetimef}
\end{equation}
The decay width to a pair of Higgs bosons is given by
\begin{equation}
\Gamma(\tp \to H \overline H)= {\delta\over n+2}\ 
{\kappa^2 m_\tp^3 \over 48\pi}(1-4r_H^{})^{1/2}(1+2 r_H^{})^2.
\end{equation}

\subsection{Effective 4-fermion Interactions}

\begin{figure}[thb]
\epsfysize=2in
\epsffile[110 360 340 480]{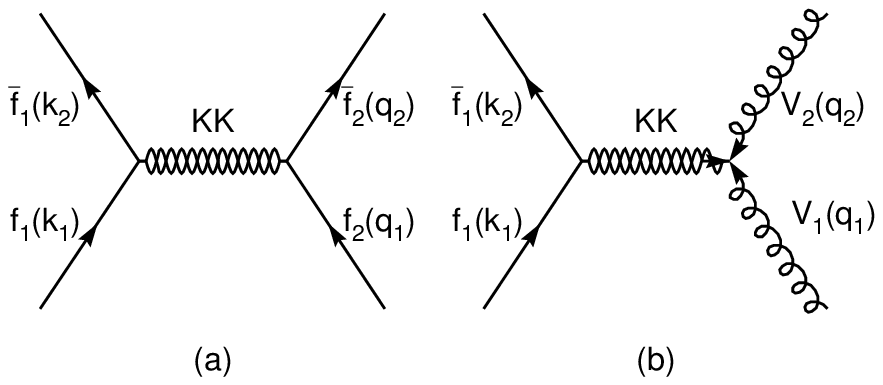}
\caption[]{Feynman diagrams for (a) four-fermion interactions and 
(b) ${\overline f}f VV$ interactions.
We represent KK states by double-sinusoidal curves.
\label{4ferm}}
\end{figure}

The most basic contribution for KK states to current 
high energy phenomenology would be the effects on 
four-fermion interactions. Consider a generic four-fermion
process
\begin{equation}
f_1(k_1)\  \bar f_1(k_2)\to  f_2(q_1)\  \bar f_2(q_2)
\end{equation}
in Fig.~\ref{4ferm}a,
where the fermion momenta are chosen to be along the 
fermion line direction. The effective amplitudes are 
calculated to have the forms
\begin{eqnarray}
\label{hfermion4}
{\cal M}_4(\th) &=& -{\pi C_4\over 2}\biggl[ (k_1+k_2)\cdot (q_1+q_2)\ 
{\overline f_2}\gamma^\mu f_2\ {\overline f_1}\gamma_\mu f_1\nonumber\\
&&+\ {\overline f_2}(\slashchar{k}_1+\slashchar{k}_2)f_2 \ 
{\overline f_1}(\slashchar{q}_1+\slashchar{q}_2) f_1
-{8\over3}m_{f_1}m_{f_2}\ {\overline f_2} f_2\ {\overline f_1} f_1 \biggr],\\
{\cal M}_4(\tp) &=&-\left({n-1\over n+2}\right){4\pi C_4\over3} m_{f_1}m_{f_2}
\ {\overline f_2} f_2\ {\overline f_1} f_1\ ,
\label{pfermion4}
\end{eqnarray}
where 
\begin{equation}
C_4 = {\kappa^2\over 8\pi} D(s),
\label{C4}
\end{equation}
and $s=(k_1-k_2)^2=(q_2-q_1)^2$.
The function $D(s)$ counts for the exchange of virtual KK states.
In principle, all the contributing KK modes in a tower should 
be summed coherently. However, the summation would be 
ultravioletly divergent for $n\ge 2$. 
We have chosen to introduce an explicit 
cutoff $M_S$ in the summation.
The full derivation and expression of $D(s)$ is given
in Appendix B. We define the relation among the gravitional 
coupling, the volume of the extra dimensions, and the cuttoff
scale as \cite{factor2}
\begin{equation}
\kappa^2 R^n= 8\pi (4\pi)^{n/2}\Gamma(n/2)M_S^{-(n+2)}.
\label{mass}
\end{equation}
Taking the leading contribution in $M_S\gg s$,
the coefficient $C_4$ then reads
\begin{eqnarray}
C_4 &\approx& -i M_S^{-4}\log(M_S^2/s)\  \quad (n=2),\\ 
 &\approx& {-2i M_S^{-4}\over (n-2)}\ 
\quad (n>2).
\end{eqnarray}
We see that the amplitude has the dimensionful pre-factor 
$M_S^{-4}$, instead of the Planck mass suppression. 
We also note that $C_4$ remains the same with $s\to |t|$ or
$|u|$ for $t,\ u$ channels. 
Thus Eqs.~(\ref{hfermion4}) and (\ref{pfermion4})
are indeed the appropriate low energy effective Lagrangians.
On the other hand, if the cutoff scale is not too far away 
from the c.~m.~energy $\sqrt s$, then the resonant contribution
in the $s$-channel should be included, as given by the
real part in Eq.~(\ref{DS}) of Appendix B.

These interactions would lead to modifications to decays of
quarkonia via
\begin{eqnarray}
(q\bar q) \to \ell\bar \ell,\  m\bar m ,
\label{decay}
\end{eqnarray}
where $(q\bar q)$ denotes a quarkonium such as
$\Upsilon,  J/\psi,\phi^0, \pi^0, \rho^0$ etc.,
$\ell=e,\mu,\tau$ and $m\bar m$ are light meson pairs.
They would also modify the scattering cross sections such as
\begin{eqnarray}
e^+e^-  &\to& \ell \bar \ell,\ q \bar q\\ 
q\bar q &\to& \ell \bar \ell,\ q \bar q.
\label{scatter}
\end{eqnarray}
Due to the particular structure of the contact interactions in 
Eq.~(\ref{hfermion4}), analyses on the final state angular 
distributions may reveal deviations from the Standard Model
predictions.

\subsection{Effective $\overline{f} f\ VV$ Interactions}

Exchanges of virtual KK states can also contribute to
processes like
\begin{equation}
f_1(k_1)\  \bar f_1(k_2)\to  V_1(q_1)\  \overline V_2(q_2),
\end{equation}
as in Fig. \ref{4ferm}b,
where the fermion momenta are chosen to be along the 
fermion line direction, and the gauge boson momenta
are incoming to the vertex. 
The effective amplitudes for fermion-gauge 
bosons should have the general form of
\begin{eqnarray}
{\cal M}_V(\th) &=& - 2\pi C_4 
\biggl[2 m_f\ (q_1\cdot V_2)\ (q_2\cdot V_1)\ {\overline f}f
+({4\over3} m_V^2 m_f - s m_f)\ (V_1\cdot V_2)\ {\overline f}f \nonumber\\
&& +2 (k_1\cdot q_2-k_1\cdot q_1)\ (V_1\cdot V_2) \ 
{\overline f} {\slashchar q}_1 f
+2 (k_1\cdot V_1) (q_1\cdot V_2)\ {\overline f} {\slashchar q}_1 f
\nonumber\\
&&-2 (k_1\cdot V_2) (q_2\cdot V_1)\ {\overline f} {\slashchar q}_1 f
-2 (k_1\cdot q_2) (q_1\cdot V_2)\ {\overline f} {\slashchar V}_1 f
+ s (k_1\cdot V_2)\ {\overline f} {\slashchar V}_1 f
\nonumber\\
&&-2 (k_1\cdot q_1) (q_2\cdot V_1)\ {\overline f} {\slashchar V}_2 f
+ s (k_1\cdot V_1)\ {\overline f} {\slashchar V}_2 f\biggr]
\\
{\cal M}_V(\tp) &=& \left({n-1\over n+2}\right)
{8\pi  C_4\over3}\ m_V^2 m_f 
(V_1\cdot V_2)\ {\overline f} f \ ,
\label{lvv}
\end{eqnarray}
where $C_4$ is the same as in Eq.~(\ref{C4}),
$s=(q_1+q_2)^2=(k_1-k_2)^2$ and $V_1,V_2$ represent 
polarization vectors of the external gauge bosons.
Examples for the induced physical processes include
\begin{eqnarray}
e^+e^-,\ q\bar q  &\to& \gamma\gamma,\ W^+W^-,\  ZZ\ {\rm and}\ gg \\ 
\gamma\gamma,\ gg &\to& \ell \bar \ell,\ q \bar q.
\label{vv}
\end{eqnarray}

\subsection{KK State Real Emission\label{sec_em}}

\begin{figure}[thb]
\epsfysize=2in
\epsffile[130 365 340 470]{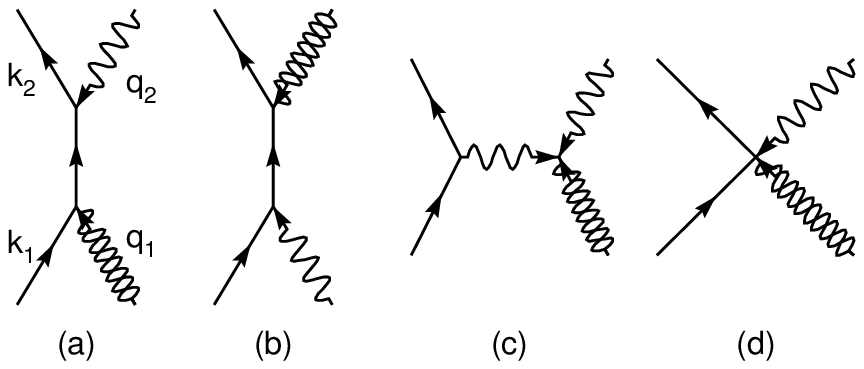}
\caption[]{Feynman diagrams for $e^-e^+\rightarrow\gamma+KK$.
\label{comp}}
\end{figure}

Since the KK states couple to all the SM particles, 
they may be radiated from quarkonium decays if kinematically
allowed, or be copiously produced at high energy colliders.
Consider the process
\begin{equation}
f\bar f \to V + KK,
\label{emission}
\end{equation}
where $V$ is a SM gauge boson.
There are four diagrams to contribute to the
process: $s,t,u$ channels plus a four-point contact diagram
as shown in Fig.~\ref{comp}.
For simplicity, we consider a massless gauge boson (a photon 
or a gluon). 

For the $\tilde \phi$ emission, it is interesting 
to note that only the fermion-mass dependent terms survive from
the $t$ and $u$ diagrams. The amplitude for the emission of
$\tilde \phi$ of mass $m_\vn$ is 
\begin{equation}
{\cal M}(\tilde \phi) = {-i\over2}\delta_{ij} \omega g_V m_f \kappa\  
{\overline u}(k_2)(
{ {\slashchar \ell}\gamma_\rho\over t} +
{ \gamma_\rho   {\slashchar j}\over u} ) u(k_1)\ \epsilon^\rho(q_2),
\label{radphi}
\end{equation}
where $\omega$ is the normalization factor in Eq.~(\ref{gen})
and $g_V=eQ_f$ for a photon and $g_sT^a_{nm}$
for a gluon, and $\ell=k_1+q_2,\ j=k_1+q_1$.
Again, our momentum convention is that the fermion momenta
follow the fermion line and the gauge boson and the KK state
have their momenta incoming to the vertices.
The amplitude for $\tilde h$ emission is calculated to be
\begin{eqnarray}
{\cal M}(\tilde h) &=& {-i\over2} g_V \kappa\ 
{\overline u}(k_2)\ \biggl[ {1\over u}\gamma_\rho {\slashchar j}
\gamma_\mu k_{1\nu} + {1\over t} \gamma_\mu k_{2\nu} 
{\slashchar \ell} \gamma_\rho + {2\over s}\gamma^\sigma
(q_1\cdot q_2 \eta_{\mu\sigma} \eta_{\nu\rho} +
\eta_{\mu\rho}k_\nu q_{2\sigma}\nonumber\\ 
& & -\eta_{\mu\sigma} q_{1\rho}q_{2\nu}
- \eta_{\rho\sigma}k_\mu q_{2\nu} )- \gamma_\mu \eta_{\nu\rho}
\biggr]\ u(k_1) \epsilon^\rho(q_2) \epsilon^{\mu\nu}(q_1),
\label{radh}
\end{eqnarray}
with $k=k_1-k_2$.
The amplitudes of Eqs.~(\ref{radphi})-(\ref{radh}) are directly
applicable to physical processes like quarkonium radiative decays
and $e^+e^-,\ q\bar q \to \gamma\ (g) + KK$, or
$e\gamma \to e + KK$ and $qg \to q + KK$. Similar calculations 
can be carried out for $W, Z + KK$ processes.

Unlike the processes with internal KK exchanges, the diagrams
for the external emission of KK modes with different masses 
do not interfere. Instead, contributions from different KK modes
will have to be summed up at the cross-section level. A general
discussion of the KK state summation is presented in Appendix B.
As an illustration, we calculate the cross-section for Eq.~(\ref{radphi}).
The cross section is given by 
\begin{equation}
\sigma = 
\left({n-1\over n+2}\right){2\pi c^2\over 3 N_c}\ {m_f^2\over s^2}\ 
(s/M_S^2)^{n/2+1}\ I_\theta\ I_y(n),
\label{realem}
\end{equation}
where $c^2=Q_f^2\alpha$ for a photon and $(N^2_c-1)\alpha_s$ 
for a gluon and $N_c$ is the number of colors. The integrals are 
\begin{equation}
I_\theta=\int_{-1+\delta}^{1-\delta}{d\cos\theta\over 1-\cos^2\theta}
=\log\left({2-\delta\over\delta}\right),\quad 
I_y(n)=\int_0^1 dy^2
{y^{n-2} (1+y^4)\over (1-y^2)^{1/2}},
\end{equation}
where $\theta$ is the photon scattering angle in the c.~m.~frame
with respect to the beam direction and $y^2= m_\vn^2/ s$.
The integral $I_\theta$ is logarithmically divergent, corresponding to
the collinear singularity ($\delta\to 0$) associated with the
massless gauge boson emission. From Eq.~(\ref{realem}), we see
once again that the cross-section rate is not suppressed by the Planck
scale rather by a power of $s/M_S^2$, due to the summation
over the large number of KK states. However, the additional
factor $m_f^2/ s$ significantly suppresses the $\tp$ emission
off light fermions. On the other hand the $\tilde h$ emission
would not have this suppression and may be phenomenologically
more interesting to study.

\subsection{One-loop Corrections from Virtual KK States}

\begin{figure}[thb]
\epsfysize=2in
\epsffile[168 368 340 450]{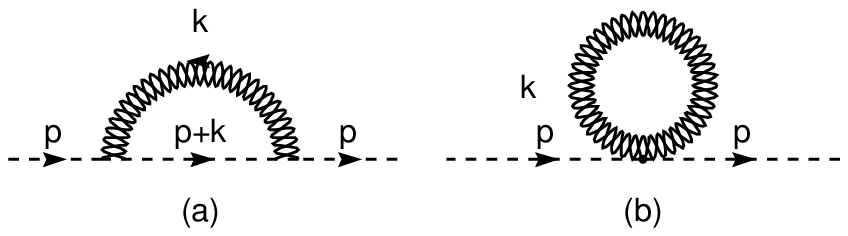}
\caption[]{One-loop self-energy diagrams of the scalar particle.
\label{sefeyn}}
\end{figure}

It is of great interest to ask what radiative effects the SM fields
may receive from the virtual KK states. 
As an example, we calculate the massive spin-2 KK state $\th^\vn_{\mu\nu}$
contribution to the one-loop self-energy for a scalar boson.
The momentum integrals involved have much worse ultraviolet 
behavior than their four-dimensional counterparts, we need to introduce
an explicit cutoff $M_S$ to regularize the ultraviolet divergence. 
 
There are two contributing diagrams, as shown in Fig.~\ref{sefeyn}.
The first one (Fig.~\ref{sefeyn}a) originates
from the KK-$\Phi\Phi$ vertex. The complete expression for this diagram
is very complicated. However, to see the leading behavior, it 
is sufficient to evaluate the self-energy at zero external momentum.
After some algebra, it can be simplified to 
\begin{equation}
-i\Pi(p=0)=i{\kappa^2\over16\pi^2}\int_0^\infty dk^2\
k^2\left({1\over k^2+m^2_\Phi}\right)
\sum_\vn\Biggl[\biggl({1\over k^2+m_\vn^2}\biggr)
\biggl(-\frac{4m^4_\Phi}{3}+{k^2m^2_\Phi}+{k^4m_\Phi^2\over m^2_\vn}
\biggr)\Biggr] ,
\label{main}
\end{equation}
where we have performed the Wick rotation. 

Since the spacing between adjacent KK states is of order ${\cal O}(1/R)$
and small, one can approximate 
the summation over the KK states by an integration, as shown in the Appendix B
\footnote{The summation over the KK states can also be calculated using
the Jacobi theta function \cite{DDG}.}. 
This reduces the above self-energy to 
\begin{equation}
-i\Pi(p=0)\ =\ {i\over 2\pi} [-{4r_\Phi\over3} I_1(n) + I_2(n)] m^2_\Phi\ ,
\end{equation}
where we have introduced an explicit ultraviolet 
cutoff $M_S$ for the momentum integration and used the relation 
Eq.~(\ref{mass}). The integrals $I_1(n)$ and $I_2(n)$ are
\begin{eqnarray}
I_1(n) &=& \int_0^1\int_0^1 dxdy\ {x y^{n/2-1}\over (x+r_\Phi)(x+y)}\ ,\\
I_2(n) &=& \int_0^1\int_0^1 dxdy\ \left({x^2 y^{n/2-2}\over x+r_\Phi}\right)\ ,
\end{eqnarray}
where $r_\Phi=m^2_\Phi/M_S^2$.

The second diagram (Fig.~\ref{sefeyn}b) comes from the
four-point KK-KK-$\Phi\Phi$ (seagull) 
vertex. To derive the Feynman rule for this vertex,
one has to expand the interaction Lagrangian to the order of $\kappa^2$.
After some tedious algebra, it can be shown that the 
Feynman rule is
\begin{equation}
i{\kappa^2\over4}\delta_{ij}
\biggl(C_{\mu\nu,\rho\sigma} m^2_\Phi
+C_{\mu\nu,\rho\sigma|\lambda\eta} k_1^\lambda k_2^\eta\biggr)\ ,
\end{equation}
where $k_1, k_2$ are four-momentum of the scalars, $C_{\mu\nu,\rho\sigma}$
is defined in Eq.~(\ref{C}) and 
\begin{equation}
C_{\mu\nu,\rho\sigma|\lambda\eta}
\ =\ {1\over2}\biggl[\eta_{\mu\lambda} C_{\rho\sigma,\nu\eta}
+\eta_{\sigma\lambda} C_{\mu\nu,\rho\eta}
+\eta_{\rho\lambda} C_{\mu\nu,\sigma\eta}
+\eta_{\nu\lambda} C_{\mu\eta,\rho\sigma}
-\eta_{\lambda\eta} C_{\mu\nu,\rho\sigma}
+(\lambda\leftrightarrow\eta)\biggr]\ .
\end{equation}

The one-loop self-energy is then
\begin{equation}
-i\Pi(p^2=-m^2_\Phi)\ =\ i{\kappa^2 m_\Phi^2\over16\pi^2}
\int_0^\infty dk^2 k^2 \sum_\vn\Biggl[\biggl({1\over k^2+m_\vn^2}\biggr)
\biggl({14\over3}+{7k^2\over3 m_\vn^2}
+{k^4\over 6m_\vn^4}\biggr)\Biggr]\ .
\label{Pi}
\end{equation}
Again we replace the summation by integration and introduce a cutoff
$M_S$, the above equation then becomes 
\begin{equation}
-i\Pi(p^2=-m^2_\Phi)\ =\ {i m^2_\Phi\over 12\pi} 
[15 I_3(n)+13 I_4(n)+I_5(n)]\ ,
\end{equation}
where 
\begin{eqnarray}
I_3(n) &=& \int_0^1\int_0^1dxdy\ \left({xy^{n/2-1}\over x+y}\right)\ ,\\
I_4(n) &=& \int_0^1dx\ x^{n/2-2}\ ,\qquad
I_5(n)\ =\ \int_0^1dx\ x^{n/2-3}\ .
\end{eqnarray}
Integrals $I_4(n)$ and $I_5(n)$ are infrared divergent
when $n\leq 2$ and 4 respectively\footnote{$I_4$ and $I_5$ come from
the summations $\sum{1\over m_\vn^2}$ and $\sum{1\over m_\vn^4}$, 
they can be regularized by the Epstein $\zeta$-function instead of 
by the explicit cutoffs.}, this is unphysical since the summation
should really start at the first nonzero mode. Therefore
a natural infrared cutoff $1/(RM_S)^2$ can be included when necessary.

It is important to note that the leading one-loop correction 
to the scalar-boson mass is proportional to $m_\Phi^2$, as opposed
to the usual cutoff ($M_S^2$) dependent corrections from other 
particles in loops.
We expect this fact to hold as well for the gauge bosons.

\section{Conclusions}

We have identified the massive KK states in the 
four-dimensional spacetime from the $(4+n)$-dimensional 
Kaluza-Klein (KK) theory, assuming 
compactification of the extra $n$ dimensions on a torus.
For a given KK level $\vec{n}$, we find that there 
are one spin-2 state, $(n-1)$ spin-1 states and 
$n(n-1)/2$ spin-0 states and they are all mass-degenerate. 

We have constructed the effective interactions among 
these KK states and ordinary matter fields (fermions, 
gauge bosons and scalars).
We find that the spin-1 states decouple and the spin-0 states
only couple through the dilaton mode.  
We derived the interacting Lagrangian for the KK states and 
Standard Model fields. These interactions are flavor-diagonal 
and thus have no new flavor-changing neutral currents, 
nor baryon and lepton number violation.
We also obtained the corresponding Feynman rules, as given 
in Appendix A, based on which further phenomenological applications
can be carried out.

For the interesting scenario when the compactification scale $1/R$ 
is small compared to experimentally accessible energies, and the cutoff 
scale is on the order of 1 TeV,
we outlined some low energy phenomenology for further studies.
Examples include quarkonium radiative decays, four-fermion 
interactions and the associated production of gauge bosons 
and KK states for those new interactions resulting from the 
massive KK modes. Although formally suppressed by the Planck
mass, the typical physical processes are only suppressed by
powers of $s/M_S^2$ after summing over the contributing KK 
states. This implies possibly significant experimental
signatures. It also recovers the ``decoupling theorem''
in the limit $M_S \to \infty$.

We also found that radiative corrections
to the scalar self-energy via virtual KK modes are proportional
to the scalar mass-squared. Finally, based on our dicussions
for the KK decays, cosmology at the early Universe should
be carefully examined with the existence of KK states in the 
extra large dimensions.

\vskip 0.1in

{\noindent\it Notes added:} 
When we are finishing this current work, another article dealing
with the same subject appeared \cite{Giudice}. 

\newpage

{\noindent\bf Acknowledgements:} We thank K. Cheung, 
B. Grzadkowski, J. Gunion, G. Landsberg, O. Lebedev 
and W. Loinaz for comments on the earlier version of this article.
T.H. and R.J.Z. were supported in part 
by a DOE grant No. DE-FG02-95ER40896 and in part by the Wisconsin
Alumni Research Foundation. J.D.L. was supported by the Fermi National 
Accelerator Laboratory, which is operated by the Universities research
Association, Inc., under contract No. DE-AC02-76CHO3000.

\vspace{0.2in}
{\noindent\Large\bf Appendix A: Feynman Rules}
\setcounter{equation}{0}
\renewcommand{\theequation}{A.\arabic{equation}}
\vspace{0.2in}

{\noindent\large\bf A.1 Propagators and Polarizations}
\vspace{0.1in}

The propagator for the massive spin-2 KK states $\th^\vn_{\mu\nu}$ 
is \cite{factor2}
\begin{equation}
i\Delta^\th_{\{\mu\nu,\vec{n}\}, \{\rho\sigma,\vec{m}\}}\ (k) 
\ =\ {i\delta_{\vec{n},-\vec{m}}\ B_{\mu\nu,\rho\sigma}(k)
\over k^2-m^2_{\vec{n}}+i\varepsilon}\ ,
\label{prop}
\end{equation}
where 
\begin{eqnarray}
B_{\mu\nu,\rho\sigma}(k) &=& 
\left(\eta_{\mu\rho}-{k_\mu k_\rho\over m_\vn^2}\right)
\left(\eta_{\nu\sigma}-{k_\nu k_\sigma\over m_\vn^2}\right)
+\left(\eta_{\mu\sigma}-{k_\mu k_\sigma\over m_\vn^2}\right)
\left(\eta_{\nu\rho}-{k_\nu k_\rho\over m_\vn^2}\right)\nonumber\\
&& - {2\over3}\left(\eta_{\mu\nu}-{k_\mu k_\nu\over m_\vn^2}\right)
\left(\eta_{\rho\sigma}-{k_\rho k_\sigma\over m_\vn^2}\right)\ .
\label{B}
\end{eqnarray}
It is obvious that $k^\mu B_{\mu\nu,\rho\sigma}=0$ and
$B^{\mu}_{~\mu,\rho\sigma}=0$ if $\th^\vn_{\mu\nu}$ 
is on shell, $k^2=m_\vn^2$.

The polarization tensors for $\th^\vn_{\mu\nu}$ can 
be constructed from the polarization
vectors of the massive vector bosons, $\epsilon_\mu^{\pm, 0}$, 
as follows:
\begin{equation}
\epsilon^s_{\mu\nu}\ =\ \Biggl\{\sqrt{2}\epsilon^+_\mu\epsilon^+_\nu,
(\epsilon^+_\mu\epsilon^0_\nu+\epsilon^0_\mu\epsilon^+_\nu),
{1\over\sqrt{3}}(\epsilon^+_\mu\epsilon^-_\nu+\epsilon^-_\mu\epsilon^+_\nu
-2\epsilon^0_\mu\epsilon^0_\nu),
(\epsilon^-_\mu\epsilon^0_\nu+\epsilon^0_\mu\epsilon^-_\nu),
\sqrt{2}\epsilon^-_\mu\epsilon^-_\nu
\Biggr\}\ .
\label{pol}
\end{equation}
These polarization tensors are traceless, 
transverse and orthogonal, 
\begin{equation}
(\epsilon^s)^{\mu}_{\ \mu}\ =\ 0\ ,\qquad 
  k^\mu\epsilon^s_{\mu\nu}\ =\ 0\ ,\qquad
\epsilon^{s, \mu\nu}\epsilon^{s'*}_{\mu\nu}\ =\ 2\delta^{ss'}\ .
\label{ortho}
\end{equation}
The completeness condition then follows from that of
$\epsilon_\mu^s$ and the definition 
Eq.~(\ref{pol}) \cite{factor2},
\begin{equation}
\sum_{s=1}^5 \epsilon^s_{\mu\nu}\epsilon^{s*}_{\rho\sigma}\ =\ 
B_{\mu\nu,\rho\sigma}(k)\ .
\label{polsum}
\end{equation}

The propagators for $\tp^\vn_{ij}$ and $\tA^\vn_{\mu i}$
have the following forms 
\begin{eqnarray}
i\Delta^\tp_{\{ij,\vn\},\{kl,\vm\}}(k) &=& 
{{i\over2}(P^\vn_{ik}P^\vn_{jl}+P^\vn_{il}P^\vn_{jk})\delta_{\vn,-\vm}
\over k^2-m_\vn^2+i\varepsilon}\ ,\\
i\Delta^\tA_{\{\mu i, \vn\},\{\nu j,\vm\}}(k) &=& 
- {i P^\vn_{ij} \delta_{\vn,-\vm}(\eta_{\mu\nu} -{k_\mu k_\nu/m_\vn^2})
\over k^2-m_\vn^2+i\varepsilon}\ ,
\end{eqnarray}
where $P^\vn_{ij}$ are the projectors defined in Eq.~(\ref{proj}).
Their appearance can be understood from the fact that
$\tp^\vn_{ij}$ and $\tA^\vn_{\mu i}$ only couple to the sources which
are dressed up by the projectors.

Since $\tA^\vn_{\mu i}$ and $\tp^\vn_{ij}$ satisfy
the divergencelessness condition in Eq.~(\ref{constraint2}),
each external state of these particles should be accompanied by 
an extra-dimension ``polarization'' vector ($e_i$) 
or tensor ($e_{ij}$), which satisfies
\begin{eqnarray}
&&
n_i e_i^s\ =\ 0\ ,~~
e_i^s e_i^{s'*}\ =\ \delta^{ss'}\ ,~~
\sum_{s=1}^{n-1} e_i^s e_j^{s*}\ =\ P^\vn_{ij}\ ,\\
&&
n_i e_{ij}^s\ =\ 0\ ,~~
e_{ij}^s e_{ij}^{s'*}\ =\ \delta^{ss'}\ ,~~
\sum_{s=1}^{n(n-1)/2} e_{ij}^s e_{kl}^{s*}
\ =\ {1\over2}P^\vn_{ik}P^\vn_{jl}+{1\over2}P^\vn_{il}P^\vn_{jk}\ ,
\end{eqnarray}
for each KK level. 

\vspace{0.2in}
{\noindent\large\bf A.2 Vertex Feynman Rules}
\vspace{0.1in}

\begin{figure}[thb]
\epsfysize=5in
\epsffile[90 150 340 500]{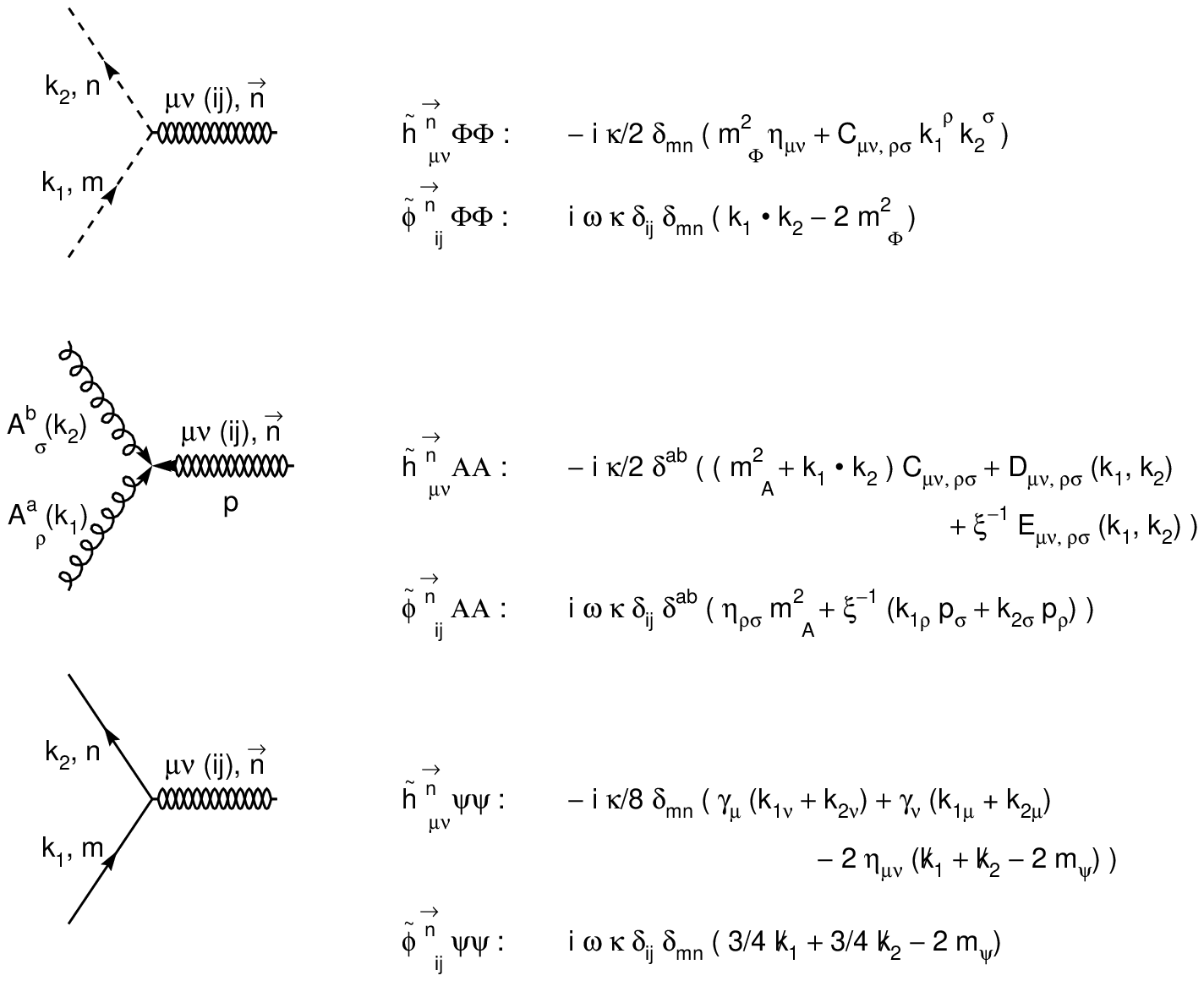}
\caption[]{Three-point vertex Feynman rules.
The KK states are plot in double-sinusoidal curves. 
The symbols $C_{\mu\nu,\rho\sigma}$, $D_{\mu\nu,\rho\sigma}(k_1,k_2)$ 
and $E_{\mu\nu,\rho\sigma}(k_1,k_2)$ are defined in Eqs. (\ref{C}),
(\ref{D}) and (\ref{E}) respectively. $m_\Phi$, $m_A$ and $m_\psi$
are masses of the scalar, vector and fermion.
$\omega=\sqrt{2\over3(n+2)}$, $\kappa=\sqrt{16\pi G_N}$
and $\xi$ is the gauge-fixing parameter.
\label{vert}}
\end{figure}

\begin{figure}[thb]
\epsfysize=5in
\epsffile[90 150 340 500]{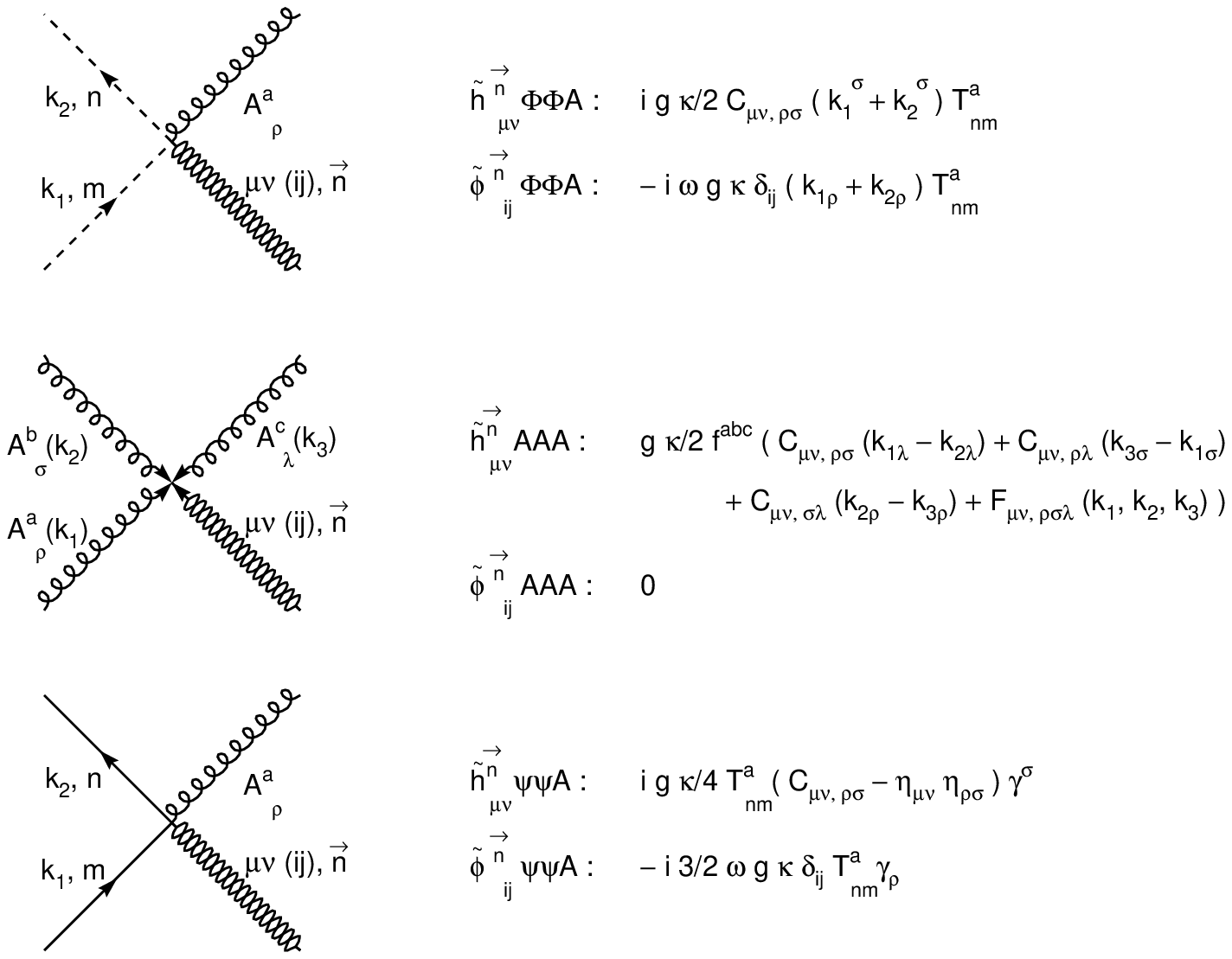}
\caption[]{Four-point vertex Feynman rules.
$g$ is the gauge coupling
and $f^{abc}$ the structure constant of the Lie algebra,
$gT^a\rightarrow eQ_f$ for QED.
The symbols $C_{\mu\nu,\rho\sigma}$ 
and $F_{\mu\nu,\rho\sigma\lambda}(k_1,k_2,k_3)$ are 
defined in Eqs. (\ref{C}) and (\ref{F}).
\label{vert2}}
\end{figure}

\begin{figure}[thb]
\epsfysize=3.5in
\epsffile[90 250 340 500]{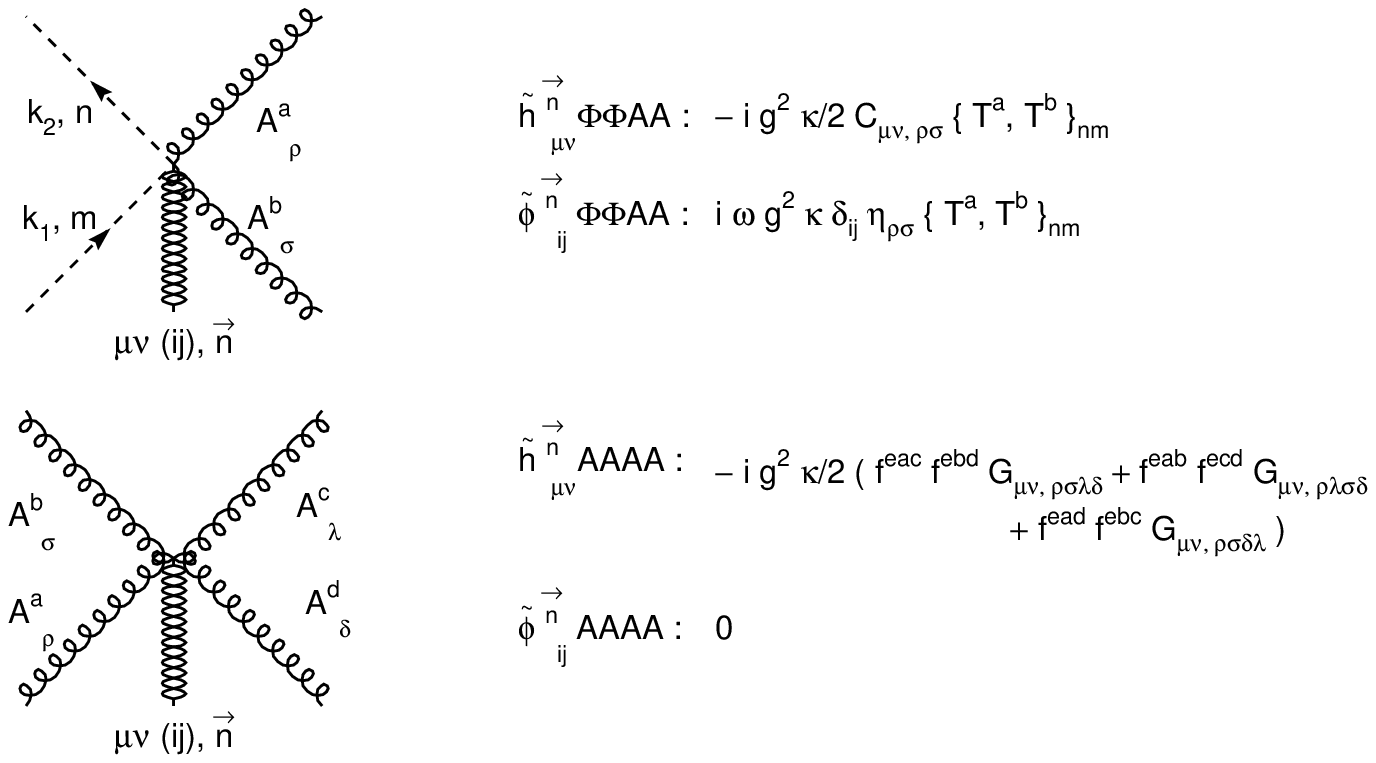}
\caption[]{Five-point vertex Feynman rules.
$g^2\{T^a, T^b\}\rightarrow 2 e^2 Q_f^2$ for QED. 
The symbols $C_{\mu\nu,\rho\sigma}$ 
and $G_{\mu\nu,\rho\sigma\lambda\delta}$ are defined in 
Eqs. (\ref{C}) and (\ref{G}).
\label{vert3}}
\end{figure}

In the following we list the complete leading order Feynman rules
in three figures, Figs. \ref{vert}, \ref{vert2} and \ref{vert3}.
Some of the symbols used are defined as follows:
\begin{eqnarray}
&&C_{\mu\nu,\rho\sigma}\ =\ \eta_{\mu\rho}\eta_{\nu\sigma}
+\eta_{\mu\sigma}\eta_{\nu\rho}
-\eta_{\mu\nu}\eta_{\rho\sigma}\ ,
\label{C}\\
&&
D_{\mu\nu,\rho\sigma} (k_1, k_2)\ =\
\eta_{\mu\nu} k_{1\sigma}k_{2\rho}
- \biggl[\eta_{\mu\sigma} k_{1\nu} k_{2\rho}
  + \eta_{\mu\rho} k_{1\sigma} k_{2\nu}
  - \eta_{\rho\sigma} k_{1\mu} k_{2\nu}
  + (\mu\leftrightarrow\nu)\biggr]\ ,\nonumber\\
&&\label{D}\\
&&
E_{\mu\nu,\rho\sigma} (k_1, k_2)\ =\ \eta_{\mu\nu}(k_{1\rho}k_{1\sigma}
+k_{2\rho}k_{2\sigma}+k_{1\rho}k_{2\sigma})
\nonumber\\
&&\qquad\qquad\qquad\qquad
-\biggl[\eta_{\nu\sigma}k_{1\mu}k_{1\rho}
+\eta_{\nu\rho}k_{2\mu}k_{2\sigma}
+(\mu\leftrightarrow\nu)\biggr]\ ,\label{E}\\
&& 
F_{\mu\nu,\rho\sigma\lambda} (k_1,k_2,k_3)\ =\  
\eta_{\mu\rho}\eta_{\sigma\lambda}(k_2-k_3)_\nu
+\eta_{\mu\sigma}\eta_{\rho\lambda}(k_3-k_1)_\nu\nonumber\\
&&\qquad\qquad\qquad\qquad\qquad
+\eta_{\mu\lambda}\eta_{\rho\sigma}(k_1-k_2)_\nu 
+ (\mu\leftrightarrow\nu)\ ,\label{F}\\
&& G_{\mu\nu,\rho\sigma\lambda\delta} \ =\  
\eta_{\mu\nu}(\eta_{\rho\sigma}\eta_{\lambda\delta}
-\eta_{\rho\delta}\eta_{\sigma\lambda})
+\biggl(\eta_{\mu\rho}\eta_{\nu\delta}\eta_{\lambda\sigma}
+\eta_{\mu\lambda}\eta_{\nu\sigma}\eta_{\rho\delta}\nonumber\\
&&\qquad\qquad\qquad\qquad
-\eta_{\mu\rho}\eta_{\nu\sigma}\eta_{\lambda\delta}
-\eta_{\mu\lambda}\eta_{\nu\delta}\eta_{\rho\sigma}
+(\mu\leftrightarrow\nu)\biggr)\ .
\label{G}
\end{eqnarray}
All of them are symmetric in $\mu\leftrightarrow\nu$.
$C_{\mu\nu,\rho\sigma}$ is the symbol that
appears in the massless graviton propagator
in the de Donder gauge.

\vspace{0.2in}
{\noindent\Large\bf Appendix B. Summation of the KK States}
\setcounter{equation}{0}
\renewcommand{\theequation}{B.\arabic{equation}}
\vspace{0.2in}

Since the KK states are nearly degenerate in mass, one would
encounter the summation over those modes that are contributing
to a given physical process. Consider the number of KK states
within a mass scale $m_\vn^2$. This is equivalent to counting
the $n$-dimensional hyper-cubic lattice sites in $\vn=(n_1,n_2...,n_n)$
with a relation to the mass
\begin{equation}
m_\vn^2 = {4\pi^2 \vn^2\over R^2},\quad {\rm or}\ ~~
r^2\equiv \vn^2= {m_\vn^2 R^2 \over 4\pi^2}.
\end{equation}
Since the mass separation of ${\cal O}(1/R)$ is much smaller 
than any other physical scale involved in the problem, it is much more
convenient to consider the discrete $\vn$ in the continuum 
limit. Therefore, the number of states in the mass interval
$dm_\vn^2$ can be obtained by
\begin{equation}
\Delta \vn^2 \approx d^nr= \rho(m_\vn)dm_\vn^2,
\end{equation}
where the KK state density as a function of $m_\vn$ is given by
\begin{equation}
\rho(m_\vn)={R^n\  m_\vn^{n-2} \over (4\pi)^{n/2}\  \Gamma(n/2)}.
\label{sph}
\end{equation}
This is the state density function that is to be convoluted
with a physical amplitude or cross-section for a KK state
with a given mass $m_\vn$.

A less trivial example is when constructing the effective 
interactions due to virtual KK state exchanges, one has to 
sum over them in the propagator
\begin{equation}
D(s) \ =\ \sum_\vn {i\over s-m_\vn^2+i\varepsilon}\ 
= \int_0^\infty dm_\vn^2 \ \rho(m_\vn) 
{i\over s - m_\vn^2 +i\varepsilon}\ ,
\end{equation}
which may be singular near a real KK state production.
Using 
\begin{equation}
{1\over s - m^2 +i\varepsilon}\ =\ P\left({1\over s - m^2}\right)
-i\pi\delta (s-m^2)\ ,
\end{equation}
we find 
\begin{equation}
D(s)\ =\ {s^{n/2-1}\over\Gamma(n/2)} {R^n\over(4\pi)^{n/2}}
\biggl[\pi + 2i I(M_S/\sqrt{s})\biggr]\ ,
\label{DS}
\end{equation}
where 
\begin{equation}
I(M_S/\sqrt{s})\ =\ P \int_0^{M_S/\sqrt{s}}dy\ {y^{n-1}\over 1-y^2}\ .
\label{B6}
\end{equation}
We have introduced an explicit ultraviolet 
cutoff $M_S/\sqrt{s}$ in the integral.
It should be understood that a point $y=1$ has been removed from the 
integration path.

The real part proportional to $\pi$ in Eq.~(\ref{DS}) is from the
narrow resonant production of a single KK mode with $m_\vn^2=s$ and the 
imaginary part $I(M_S/\sqrt{s})$ is from the
summation over the many non-resonant states. 
This principal integration of Eq.~(\ref{B6}) 
can be easily carried out, it gives
\begin{eqnarray} 
I(M_S/\sqrt{s})\ &=&\ 
-\sum_{k=1}^{n/2-1} {1\over2k}\left({M_S\over\sqrt{s}}\right)^{2k}
-{1\over2}\log\left({M_S^2\over s}-1\right)\qquad 
n={\rm even},\\ 
{} &=&\ 
-\sum_{k=1}^{(n-1)/2} {1\over2k-1}\left({M_S\over\sqrt{s}}\right)^{2k-1}
+{1\over2}\log\left({M_S+\sqrt{s}\over M_S-\sqrt{s}}\right)\quad 
n={\rm odd}.\nonumber
\end{eqnarray}
For $M_S \gg \sqrt s$, the leading contribution comes from the 
non-resonant states and yields
\begin{eqnarray}
D(s)&\approx& {-i\over4\pi} R^2 \log(M_S^2/s)\  \quad (n=2),\nonumber\\ 
&\approx& {-2i\over(n-2)\Gamma(n/2)} 
{R^n  M_S^{(n-2)}\over(4\pi)^{n/2}}\ \quad (n>2).
\label{DSa}
\end{eqnarray}

The summation of space-like propagators can be evaluated
similarly, and it gives
\begin{equation}
D_E(t)\ =\ \sum_\vn {i\over t-m_\vn^2}\ =\  \sum_\vn {-i\over |t|+m_\vn^2}\
=\ {|t|^{n/2-1}\over\Gamma(n/2)} {R^n\over(4\pi)^{n/2}}\ (-2i)
I_E(M_S/\sqrt{|t|}\ )\ ,
\end{equation}
where the integral $I_E$ is
\begin{eqnarray}
&&I_E(M_S/\sqrt{|t|})\ =\ \int_0^{M_S/\sqrt{|t|}}dy\ {y^{n-1}\over 1+y^2}
\nonumber\\
&=& 
(-)^{n/2+1}\Biggl[
\sum_{k=1}^{n/2-1} {(-)^{k}\over2k}\left({M_S\over\sqrt{|t|}}\right)^{2k}
+{1\over2}\log\left({M_S^2\over |t|}+1\right)\Biggr]\ 
\qquad n={\rm even}\ ,\\ 
&=&
(-)^{(n-1)/2}\Biggl[\sum_{k=1}^{(n-1)/2} {(-)^{k}\over2k-1}
\left({M_S\over\sqrt{|t|}}\right)^{2k-1}
+\tan^{-1}\left({M_S\over \sqrt{|t|}}\right)\Biggr]\ 
\qquad n={\rm odd}\ .\nonumber
\end{eqnarray}
We note that leading terms in $D_E(t)$ 
for $M_S^2\gg |t|$ are exactly of the same form as in Eq.~(\ref{DSa})
and lead to $D_E(t)=D(s\to |t|)$. This shows that the low energy 
effective interactions for $s$ and $t$ channels are equivalent.

\bigskip

\bibliographystyle{unsrt}

\begin{thebibliography}{99}
\bibitem{factor2}
In the earlier version of this article,
there was an incorrect factor of $1/2$ in Eqs.~(\ref{prop}) 
and (\ref{polsum}). With the new normalization ($8\pi$)
in Eq.~(\ref{mass}),
the physical cross sections for the spin-2 KK gravitons
remain the same; while those for the scalar dilatons are
reduced by 2.
\end{thebibliography}

\end{document}